\documentclass[prd,twocolumn,superscriptaddress,altaffilletter,amssymb,showpacs,epsfig,nofootinbib]{revtex4}
\usepackage[dvips]{graphicx}
\usepackage{amsmath}
\newcommand{\be}{\begin{equation}}
\newcommand{\ee}{\end{equation}}
\newcommand{\ben}{\begin{eqnarray}}
\newcommand{\een}{\end{eqnarray}}
\newcommand{\vphi}{\varphi}
\newcommand{\ds}{\displaystyle}

\begin{document}

\title{Quintom cosmologies admitting either tracking or phantom attractors}
\author{Ruth Lazkoz}
\email{ruth.lazkoz@ehu.es}
\affiliation{Fisika Teorikoa, Zientzia eta Teknologia Fakultatea, Euskal Herriko Unibertsitatea, 644 Posta Kutxatila, 48080 Bilbao, Spain}
\author{Genly Le\'on}\email{genly@uclv.edu.cu}\affiliation{Universidad Central de Las Villas, Santa Clara CP 54830, Cuba}
\date{\today}

\begin{abstract}
In this paper we investigate the evolution of a class of  cosmologies  fuelled by quintom dark energy  and dark matter. Quintom dark energy is a hybrid of quintessence and phantom which involves the participation of two reals scalar fields playing the roles of those two types of dark energy. In that framework we examine
from a dynamical systems perspective the possibility that those fields are coupled among them by considering an exponential potential with an interesting  functional dependence similar but not identical to others studied before. The model we consider represents a counterexample to the typical
behavior of quintom models with exponential potentials because it admits either tracking  attractors ($w=0$), or phantom attractors ($w<-1$).
\end{abstract}

\maketitle

\section{Introduction}
Understanding the nature of dark energy has become a joint non trivial pursuit of an everyday growing number of cosmologists. It is expected that the large investment which is being made in technical and human resources will pay off in the form of a deeper understanding of the link between the physics governing the universe at the classical and quantum levels.

Abundant observational evidence  supports the view that the universe is currently  undergoing accelerated expansion \cite{obs1, obs2, obs3, obs4, obs5, obs6} due to a time-evolving component of unknown nature called dark energy. Peculiarly, in most of the analyses the best agreement with observational data is provided by models in which, as time goes forward, the dark energy equation of state parameter $w=p/\rho$ crosses the $w=-1$ divide (to become even more negative). However, among recent studies using newest SN data, some suggest the convenience of the phantom divide crossing may just be an illusion due to systematic errors in observations \cite{vanish}, whereas some others conclude that all classes of dark energy models are comfortably allowed \cite{barger}.  The  number of studies with a negative regard on the existence of a $w<-1$ epoch are still scarce, and to add to the problem results have strong dependence on the chosen parametrization of $w$ in terms of redshift \cite{doran}, so there is not a full consensus about the observational preference of the phantom divide crossing. For this reasons, we feel much has to be done yet before one can definitely say the scale  tilted to one side, so continuing to get deeper insight into  models which do the crossing seems still justified.

Some interesting general aspects of the problem of the phantom divide crossing were discussed in \cite{caldor}, where the viability requirements on the equation of state and sound speed were analyzed. 
Even though, some of realizations of the crossing
can have an extradimensional origin, either in the brane \cite{brane} or the string gas context \cite{string}, scalar field models in standard four dimensional physics 
 are the most popular options of the inventory. The 
 impossibility of the occurrence of the transition in traditional single field models \cite{theoretical} has motivated much activity in the construction of two field models that do the job. 
Examples of explicit constructions can be found in \cite{hu,so,hybrid}, but perhaps the class of models which have received most attention are quintom cosmologies \cite{quintom,hess,per, dscross1, dscross2,hybrid}. 

Other worth mentioning alternatives are models in the framework of scalar-tensor theories \cite{scalar}, a single field proposal involving high order derivative operators in the lagrangian \cite{high},  a model with a single dynamical scalar field coupled to an a priori
non-dynamical background covariant vector field \cite{vector}, and an interacting Chaplygin gas \cite{chap}. We stick here to the main stream and consider quintom cosmologies as well.

The cosmological evolution of quintom models has been examined using standard dynamical systems techniques in \cite{dscross1} and \cite{dscross2}. In these references an isotropic and homogeneous universe was supposed to be filled with dark matter and dark energy, and the potential of the latter was chosen to be of exponential form. The difference between \cite{dscross1} and \cite{dscross2} is that in the second case the potential accounts for an interaction  between the conventional and the phantom scalar fields. Here we revisit quintom models from the dynamical systems perspective, but make a choice of potential which is closely related to that in \cite{dscross2}, but which at the same time does not include it as a particular case. The subtle difference of our choice has  unexpected consequences. Firstly, there exists the possibility that quintom dark energy (with fairly general initial conditions) tracks dark matter, thus meaning that avoidance of tracking behavior in quintom models is not generic. Secondly, in some cases for which the parameters in the potential exclude the tracking attractor, then the model may  have a purely phantom attractor, i.e. this will be models in which $w$ relaxes to be less than $-1$. This is the converse of the behavior observed in previously studied models with exponential potentials, in which the late-time asymptotic behavior corresponded simply to $w=-1$, despite the presence of  a transitory epoch with $w<-1$. In our opinion, this finding shows the casuistics of the evolutionary behavior of quintom dark energy models is in fact reacher than was known up to the date, and makes it a subject of study worth of further investigations. 

\section{The model}
We investigate the evolution of a spatially flat Friedmann-
Robertson-Walker (FRW) spacetime filled with dark matter with energy density $\rho_m$ and quintom dark energy with energy density $\rho_{de}$ and pressure $p_{de}$.

Our model is assumed to obey the standard Friedmann equation and energy conservation equations, that is,
\begin{eqnarray}
&&3H^2=\rho_m+\rho_{de}\,,\label{F}\\
&&\dot\rho_{de}+3H(\rho_{de}+p_{de})=0\,,\label{consde}\\
&&\dot\rho_m+3H\rho_m=0\,\label{consm}.
\end{eqnarray}
Here and throughout overdots denote differentiation with
respect to cosmic time $t$, $H=\dot a/a$ is the Hubble factor, and 
$a$ is the synchronous scale factor.

Combining Eqs. (\ref{F},\ref{consde}, \ref{consm}) one obtains the evolution equation of $H$, i.e.,
\begin{eqnarray}
-2\dot H&=\rho_m+\rho_{de}+p_{de}\label{ray}.
\end{eqnarray}

Dynamical systems are a well-known tool for depicting  the asymptotic (and sometimes the intermediate) behavior of cosmological models  by making use of the concepts of past and future attractors. We, thus, take advantage of that fact and apply it to the model under consideration. One of the first tasks involved in this protocol is introducing a set of convenient (expansion normalized) variables which allow  casting the conservation equations and the evolution equation of $H$ as a dynamical system,
subject to a constraint arising from the Friedmann equation (\ref{F}).

In our quintom cosmologies a minimum of four variables are needed to construct
an autonomous dynamical system, but because of the existence of the constraint one can always ``forget about'' the evolution of one of the variables. Let us consider just for a while we are using $n$ variables which we will denote as $x^i$.  The equations will be of the form
\be
{x^i}\hspace{1pt}'=f^i(x^1,x^2,\dots,x^n).
\ee The prime are used to denote differentiation with respect to an alternative new time variable $\tau$ which has to be chosen upon convenience dictated by the normalization of the variables (here a very standard choice will suffice).

\begin{table*}[t!]\caption[crit]{Location, existence and decelation factor of the critical points for $\bar m>0$,  $\bar n>0$ and $y>0$.}
\begin{tabular}{@{\hspace{4pt}}c@{\hspace{14pt}}c@{\hspace{14pt}}c@{\hspace{14pt}}c@{\hspace{18pt}}c@{\hspace{18pt}}c@{\hspace{2pt}}}
\hline
\hline\\[-0.3cm]
Name &$x_\phi$&$x_\vphi$&$y$&Existence&$q$\\[0.1cm]
\hline\\[-0.2cm]
$O$& $0$& $0$& $0$& All $\bar m$ and $n$ &$\ds\frac{1}{2}$\\[0.2cm]
$C_{\pm}$ & $\pm\sqrt{1+{x_{\varphi}^*}^2}$& $x_{\varphi}^*$& $0$&
All $\bar m$ and $n$ &$2$ \\[0.2cm]
${P}$&
$\tilde{m}$& $-\tilde{n}$ &$\sqrt{1-\tilde{m}^2+\tilde{n}^2}$&$\bar m^2-\bar n^2< 1$&$-1+3(\tilde{m}^2-\tilde{n}^2)$\\[0.2cm]
${T}$&
$\ds\frac{\ds\tilde{m}}{\ds2(\tilde{m}^2-\tilde{n}^2)}$&
$-\ds\frac{\ds\tilde{n}}{\ds2(\tilde{m}^2-\tilde{n}^2)}$&
$\ds\frac{1}{2\sqrt{\tilde{m}^2-\tilde{n}^2}}$&$\bar m^2-\bar n^2\ge1/2$&$\ds\frac{1}{2}$\\[0.4cm]
\hline
\hline
\end{tabular}\label{tab1}
\end{table*}

The next step in the study of the evolution of our dynamical system  is to find its fixed (or critical) points $(x^{1\star},x^{2\star},\dots, x^{n\star})$, which are given by the conditions
\begin{equation}
f^i(x^{1\star},x^{2\star},\dots, x^{n\star})=0.
\end{equation}
 The stability of the fixed points $(x^{1\star},x^{2\star},\dots, x^{n\star})$ is then analyzed by studying the  linearized dynamical system obtained by expanding the $n$ evolution equations 
about those fixed points (as explained in many seminal references, e.g \cite{hirsch}). 
Afterward, one tries solutions in the form $(x^1,x^2,\dots,x^n)=(c_1,c_2,\dots,c_n)\,e^{\lambda t}$ in the linear
approximation, and finds that their characteristic exponent $\lambda$ and the constant
vector $(c_1,c_2,\dots,c_n)$ must be respectively an eigenvalue and an eigenvector of the matrix

\begin{equation}\\
\left( \begin{array}{cccc}
\displaystyle\frac{\partial {x^1}\hspace{1pt}'}{\partial x^1}&   
\displaystyle\frac{\partial {x^1}\hspace{1pt}'}{\partial x^2}&   
\dots&
\displaystyle\frac{\partial {x^1}\hspace{1pt}'}{\partial x^n}\\
\vspace{-8pt}\\
   \displaystyle\frac{\partial {x^2}\hspace{1pt}'}{\partial x^1}&   
\displaystyle\frac{\partial {x^2}\hspace{1pt}'}{\partial x^2}&   
\dots&
\displaystyle\frac{\partial {x^2}\hspace{1pt}'}{\partial x^n}\\
\vspace{-8pt}\\
\vdots&\vdots&\dots&\vdots\\
\displaystyle\frac{\partial {x^n}\hspace{1pt}'}{\partial x^1}&\displaystyle\frac{\partial {x^n}\hspace{1pt}'}{\partial x^2}&\dots&\displaystyle\frac{\partial {x^n}\hspace{1pt}'}{\partial x^n}
\end{array} \right)_{(x^{1},x^{2},\dots, x^{n})=(x^{1\star},x^{2\star},\dots, x^{n\star})}\\
.\vspace{5pt}\\
\end{equation}

Let us know describe the  basics of the character of the fixed points depending on the values of the  characteristic exponents. 
If the real part of the three characteristic exponents is negative, the
fixed point is asymptotically stable, i.e., an attractor. On the other
hand, it is enough to have (at least) one characteristic exponent with
positive real part to make the fixed point asymptotically unstable, i.e., a repeller. This
repeller is a saddle point if at least one of the other characteristic
exponents has a negative real part, in which case there is, apart from the
unstable manifold, a stable manifold containing the exceptional orbits
that converge to the fixed point. The case in which the largest real
part is precisely zero must be analyzed  using other methods, because the
linear analysis is inconclusive. The geometric form of the orbits near
the fixed point is determined by the imaginary part of the
characteristic exponents. If the three are real the fixed point is a
node. A couple of complex conjugate exponents leads, except in
degenerate cases, to a spiral saddle  (the orbits are helices near the
fixed point). In addition, when one of the exponents is null the point is not hyperbolic
and therefore structural stability cannot be guaranteed (the geometric form of the trajectories may change under small perturbations).

In the next sections we are going to present our quintom model and will study its  evolution equations. We will find and characterize the fixed points and we will analyze the cosmological consequences of the results obtained. Special attention will be paid to the possibility these models offer to explain the passage from a conventional to a phantom equation of state. 

\section{Phase-space}

We begin by presenting Eqs. (\ref{consde},\ref{consm},\ref{ray}) in the form of a dynamical system. This is done  by making the following convenient choice of variables:
\begin{equation}
x_\phi=\frac{\dot \phi}{\sqrt{6}H},\;x_\vphi=\frac{\dot \vphi}{\sqrt{6}H},\; y=\frac{\sqrt{V}}{\sqrt{3}H},\;z=\frac{ \sqrt{\rho_m}}{\sqrt{3}H}\label{vars}.
\end{equation}
which renders the Friedmann equation as
\begin{eqnarray}
x_\phi^2-x_\vphi^2+y^2+z^2=1.   
   \label{ct}
   \end{eqnarray}

Eq. (\ref{F}) leads to the constraint (\ref{ct}) which allows considering only the evolution of those three variables solely (the evolution of the fourth one been determined by the evolution of the former).  In this case we will choose $z$ as the  variable one can mostly do without. 

The choice of potential is crucial to our discussion, we will set
\be
V=V_0\,e^{-\sqrt{6}(\bar m\phi+\bar n\varphi)},
\ee
with $\bar m$, $\bar n$ and $V_0$ positive  constants.  This kind of potential was considered to account for the interaction of conventional scalar fields for instance in \cite{copeland}.

In this paper we will only be concerned with expanding universes, that is, in what follows we will restrict ourselves to $H\ge0$ or equivalently to $y\ge0$. Combining expressions (\ref{F},\ref{consde},\ref{consm},\ref{ray},\ref{vars}), the following evolution equations are obtained:
\ben
&&x_\phi'=\frac{1}{3} \left(3 \bar m y^2+(q-2) x_\phi\right)\\
&&x_\vphi'=-\frac{1}{3} \left(3 \bar m y^2-(q-2) x_\varphi \right)\\
&&y'=\frac{1}{3} (1+q-3(\bar m x_\phi+\bar n x_\varphi)) y
\een

Here primes denote differentiation with respect to a new time variable $\tau=\log a^3$ and $q\equiv-\ddot a a/\dot a^2$ stands for the deceleration factor. Explicitly
\be
q=\frac{1}{2} \left(3 \left(x_\phi^2-x_\varphi^2-y^2\right)+1\right).
\ee

\begin{figure}[t!]
\begin{center}
\hspace{0.4cm}
\includegraphics[width=6.5cm,height=5cm]{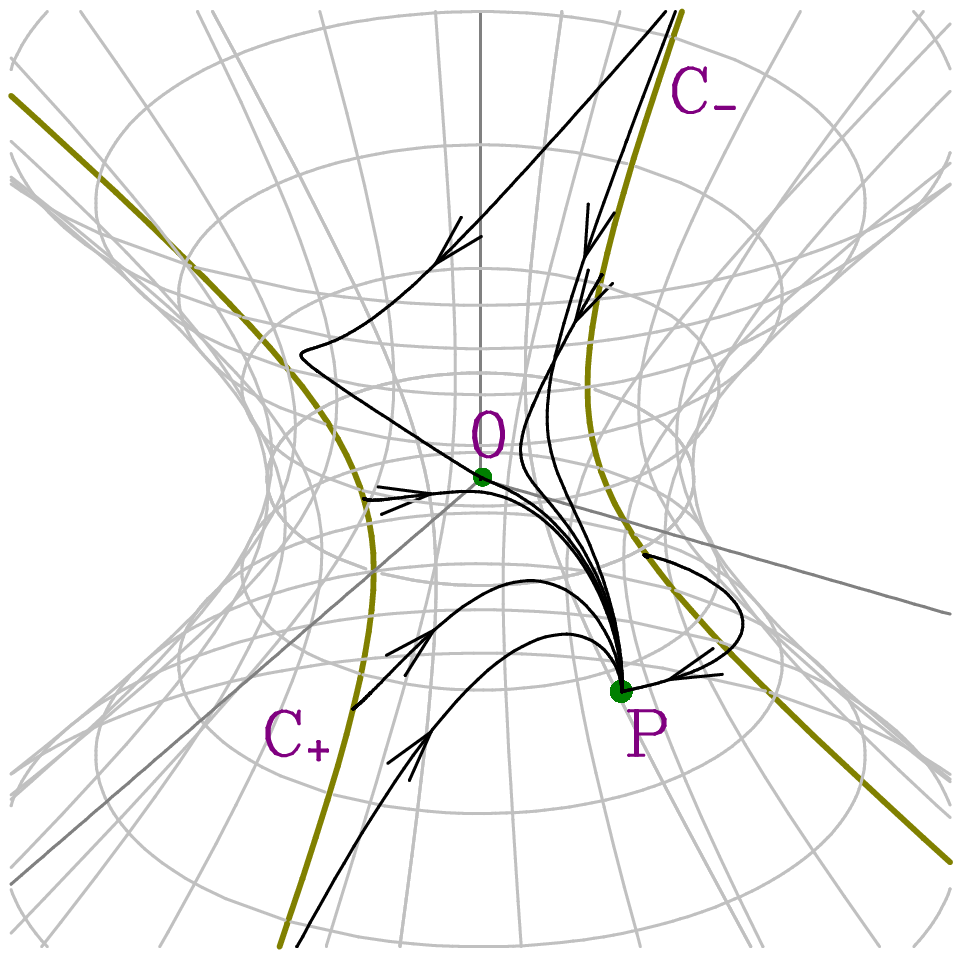}
\put(-180,5){$x_{\phi}$}
\put(-90,130){$x_{\vphi}$}
\put(-10,58){$y$}
\caption{Phase-space trajectories of the 3D system for $\bar m=0.5$ and $\bar n=0.6$. }
\label{noT}
\end{center}\end{figure}

\begin{figure}[t!]
\begin{center}
\hspace{0.4cm}
\includegraphics[width=6.5cm,height=5cm]{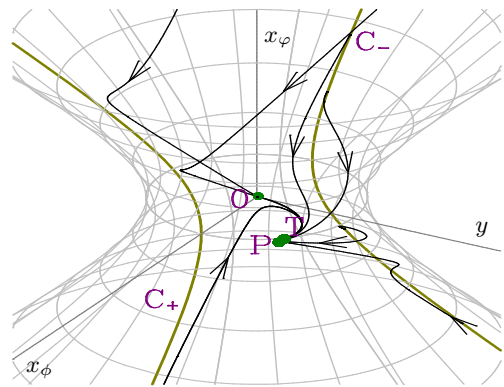}
\put(-180,5){$x_{\phi}$}
\put(-90,130){$x_{\vphi}$}
\put(-10,58){$y$}
\caption{Phase-space trajectories of the 3D system for $\bar m=0.75$ and $\bar n=0.05$. }
\label{TnodePsaddle}
\end{center}
\end{figure}

The evolution equations of variables $x_{\phi}$, $x_{\vphi}$ and $y$ form a 3D dynamical system  defined on the  phase space
\be\Psi=\{(x_\phi,x_\vphi,y):0\le x_\phi^2-x_\vphi^2+y^2\le1\}.\ee
Thus, the phase space trajectories live in an hyperboloid.

The  linear analysis described above tells us the 3D dynamical system under consideration has  at most three isolated points (depending on the values of $\bar m$ and $\bar n$). We will denote them as $O$, $T$ and $P$. In addition, there are two curves (hyperbolas) of non-isolated fixed points, and we will denote them as $C_{\pm}$.

Some information regarding the dynamical character of those fixed points is presented in
Tables (\ref{tab1},\ref{tab2}). Below we will complete that information with an identification of the cosmological models represented by the fixed points of our system, and in the case of the isolated fixed points we will also comment on their geometry. Finally, we will support our findings with numerical simulations. Note that in the case of 3D dynamical systems it is in general difficult to retrieve enough information on the system without resorting to numerical experiments \cite{reza}. The need to employ numerical tools is particularly manifest in the problem we are concerned with, 
because the asymptotic and intermediate behavior stand on the same basis of importance. Bear in mind that we do not only want to show our system admits solutions with equation of state parameter $w>-1$ and other with $w<-1$, but we also want to show that a transition between such solutions can occur without too much fine-tuning in initial conditions. In our case
\be
w=\frac{x_\phi^2-x_\vphi^2-y^2}{x_\phi^2-x_\vphi^2+y^2},
\ee
and we will evaluate the latter at the fixed points. 

\begin{table*}\caption[crit2]{Eigenvalues, dynamical character and equation of state parameter of the fixed points  for $\bar m>0$ and $\bar n>0$. We use $\Delta=\sqrt{-7+4/(\tilde{m}^2-\tilde{n}^2)}$ .\\[-0.2cm]}
\begin{tabular}{@{\hspace{4pt}}c@{\hspace{14pt}}c@{\hspace{18pt}}c@{\hspace{18pt}}c@{\hspace{2pt}}}
\hline
\hline\\[-0.3cm]
Name &Eigenvalues&Dynamical character&$w$\\[0.1cm]
\hline\\[-0.2cm]
$O$& $\ds\left(-\frac{1}{2},-\frac{1}{2},\frac{1}{2}\right)$& unstable& undefined \\[0.2cm]
$C_{\pm}$ & $\ds\left(1,0,1-\tilde{n}x_{\varphi}^*\mp
\tilde{m}\sqrt{1+{x_{\varphi}^*}^2}\right)$ & unstable & $1$ \\[0.2cm]
${P}$&$\left(-1+2(\tilde{m}^2-\tilde{n}^2),-1+\tilde{m}^2-\tilde{n}^2,
-1+\tilde{m}^2-\tilde{n}^2\right)$&
stable if $\ds \tilde m<\sqrt{\frac{1}{2}+\bar n^2}$, unstable otherwise
& $-1+2(\tilde{m}^2-\tilde{n}^2)$ \\[0.2cm]
${T}$&
$\left(-\ds\frac{1}{2},-\ds\frac{1}{4}(1+\Delta),-\ds\frac{1}{4}(1-\Delta)\right)$&
stable if either $\Delta^2<0$ or if $1>\Delta$,&$0$\\[-0.2cm]
&&
 unstable otherwise&
\\[0.4cm]
\hline
\hline
\end{tabular}\label{tab2}
\end{table*}

In broad terms, if one of the asymptotic limits of our model is to give good description of the universe at present it must be an accelerating solution. Thus, non-accelerated solutions must be asymptotically unstable, or put another way, they must be disfavored by initial conditions. In this sense only the cases in which $P$ is stable would be satisfactory. 

As can be deduced from Tables (\ref{tab1},\ref{tab2}), many of the features of our dynamical system depend on the value of the quantity $\bar m^2-\bar n^2$.  One can  easily realize why such quantity may be of  relevance just by noticing
\be
6(\bar m^2-\bar n^2)=\left(\frac{V_{,\phi}}{V}\right)^2-\left(\frac{V_{,\vphi}}{V}\right)^2,
\ee
i.e. the quantity marking the existence of one attractor or the other compares the slope of the potential in two different directions, or equivalently it compares how fast the fields release (gain) potential energy when they roll down (climb up) the potential. Unlike in single scalar field models with exponential potentials, one can have accelerated expansion even if the potential is not too shallow, what matters here is not how flat  the potential is, but rather whether it is much flatter in one direction that in the other.

\begin{figure}[t!]
\begin{center}
\hspace{0.4cm}
\includegraphics[width=6.5cm,height=5cm]{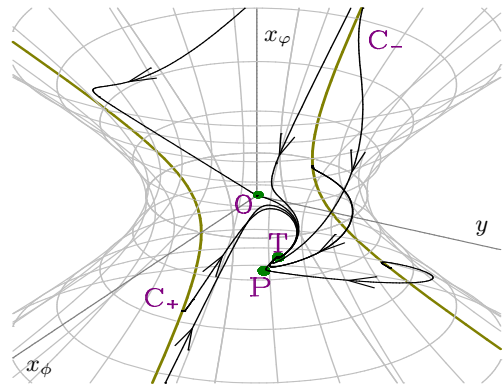}
\put(-180,5){$x_{\phi}$}
\put(-90,130){$x_{\vphi}$}
\put(-10,58){$y$}
\caption{Phase-space trajectories of the 3D system for $\bar m=0.9$ and $\bar n=0.4$. }
\label{TspiralPsaddle}
\end{center}
\end{figure}

\begin{figure}[t!]
\begin{center}
\hspace{0.4cm}
\includegraphics[width=6.5cm,height=5cm]{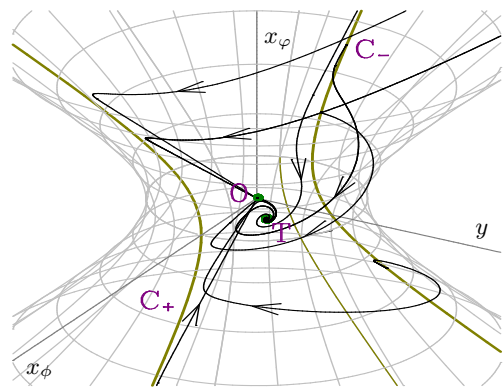}
\put(-180,5){$x_{\phi}$}
\put(-90,130){$x_{\vphi}$}
\put(-10,58){$y$}\caption{Phase-space trajectories of the 3D system for $\bar m=2$ and $\bar n=0.5$. }
\label{spiralsaddleTnoP}
\end{center}
\end{figure}

After these preliminary remarks, let us now be more specific about the dynamical character of the isolated and non-isolated fixed points of our system.

The point $O$ represents a matter dominated decelerating solution. It is a saddle, and its unstable character (see all four figures) agrees with what  one might have anticipated. Such  solutions are only
expected to be relevant at early times. The variables we are using do not allow to determine the value of equation of state parameter corresponding in this case (because we do not know the values of the different ratios between $x_\phi$, $x_\vphi$ and $y$).  This task would
possibly require defining an alternative set of coordinates more suitable for an early time description, but this is of little interest in the context of this paper.

The curves $C_\pm$ represents solutions in which the contribution of matter and the potential energy to the total  energy density is negligible. These solutions are therefore of stiff-fluid type ($w=1$), which in turn correspond to a decelerating universe. The unstable character of  this solutions means they are disfavored from the initial conditions point of view, and therefore it is unlikely they can represent the final stage in the evolution of our universe.

Further interesting remarks concerning the curves $C_\pm$ are in order to help anticipate
the intermediate behavior of our dynamical system. In the invariant set $y=0$ which contains the point $O$ as well as the curve $C_\pm$ we have identified the monotonic function ${\cal K}$, which is defined as 
\be
{\cal K}=\frac{x_\phi^4}{(1-x_\phi^2+x_\varphi^2)^2}.
\ee
It satisfies 
\be
{\cal K}'=-2{\cal K},
\ee
so as expansion proceeds ${\cal K}$ decreases. This monotonic function takes its maximum and minimum values 
at $x_\phi^2-x_\varphi^2=1$ and $x_\phi=0$, so typically trajectories will transit between
$C_\pm$ and $O$, and one will be able to say the curves $C_\pm$ are local sources. This appreciation is supported by the four figures.

The point $T$ represents a solution in which quintom dark energy tracks matter (the equation of state of the quintom fluid is dust-like), and when such point exists it is an attractor.
This late-time asymptotic state represents a decelerating cosmological model in which the fractional energy densities of matter and dark energy are proportional. Given its eigenvalues structure this solution is either a saddle or an spiral saddle. Since this solutions do not match the observed universe at present, a quintom model admitting such attractor is not satisfactory. Peculiarly, the existence of this solution is completely inherent to the interaction between the two fields. There is no equivalent solution in the quintom models studied by.   Its existence, in our setup at least, indicates quintom models admit other attractors than phantom or de Sitter ones, a fact which has not been noticed in the literature before (as far as we are concerned).

The point $P$ represents a solution in which quintom dark dominates over matter (the equation of state of the quintom fluid corresponds to a fluid that redshifts faster than dust), and when such point exists it is an attractor.
This late-time asymptotic state does not necessarily represents an accelerating cosmological model, that depends on the quantity $\bar m^2-\bar n^2$. Given its eigenvalues structure this solution is either a saddle or a stable node.  The accelerated solutions associated with this fixed point can provide a good representation of the observed universe at present. The potential we have chosen here allows this fixed point be characterized by $w<-1$, whereas in the  quintom models studied by  \cite{dscross1,dscross2}  the attractor had $w=-1$ necessarily.

\begin{figure}[t!]
\begin{center}
\hspace{0.4cm}
\includegraphics[width=9cm,height=5cm]{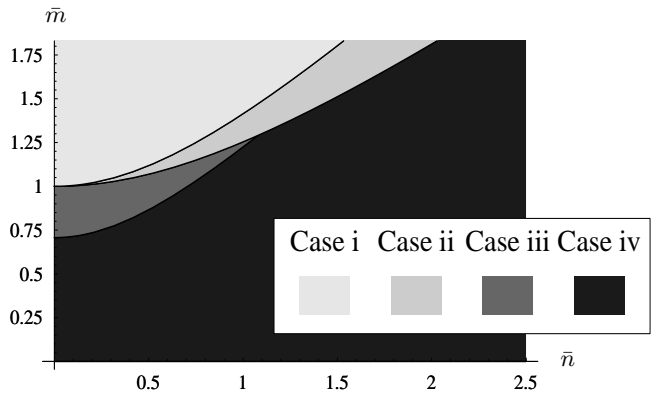}
\put(-235,145){$\bar m$}
\put(-40,15){$\bar n$}
\caption{Regions in the $(\bar n,\bar m)$ plane representing the four cases of existence conditions and dynamical behavior of the fixed points $T$ and $P$.}
\label{regions}
\end{center}\end{figure}

According to their eigenvalues, in principle the fixed points $T$ and $P$ could respectively have three and two different dynamical behaviors\footnote{The point $T$ can either be a saddle, a spiral saddle, or a stable node. The fixed point $P$ can be a saddle or a stable node.}. However, in some cases one point have a specific dynamical behavior enforces the nonexistence of the other point. Additionally, the existence conditions of these points on one hand, and numerical analysis on the other, will help us identify the heteroclinic sequences of these models. Under the assumptions $\bar m>0$ and $\bar n>0$, four cases can be distinguished (see Fig. \ref{regions}):
\begin{itemize}
    \item Case i) For $\tilde{m}<\sqrt{\tilde{n}^2+1/2},$ the point $P$ is a stable node, whereas the point $T$
    does not exist.  The heteroclinic sequence
in this case is $C_{\pm}\longrightarrow O\longrightarrow P$ (see Fig. \ref{noT}).
    \item Case ii) For $\sqrt{\tilde{n}^2+1/2}<\tilde{m}\leq
    \sqrt{\tilde{n}^2+4/7},$ the point $T$ is a stable node and the point $P$ is
a saddle. For these conditions the heteroclinic sequence is
$C_{\pm}\longrightarrow O\longrightarrow T \longrightarrow P $ (see Fig. \ref{TnodePsaddle}).
    \item Case iii) For $
    \sqrt{\tilde{n}^2+4/7}<\tilde{m}<\sqrt{1+n^2},$ the point
T is a spiral saddle and the point P is a saddle. For these conditions
the heteroclinic sequence is the same as in the former case (see Fig. \ref{TspiralPsaddle}).
    \item Case iv) For $\tilde{m}>\sqrt{1+\tilde{n}^2}$ the point T is a
spiral saddle whereas the point $P$ does not exist. The heteroclinic
sequence in this case is  $C_{\pm}\longrightarrow O\longrightarrow
T$ (see Fig. \ref{spiralsaddleTnoP}).
\end{itemize}

Before closing this section it is interesting to make a few remarks about the fractional energy densities $\Omega_m=\rho_m/{3H^2}$ and $\Omega_{de}=\rho_{de}/{3H^2}$. It is not difficult to see that
\be
\left(\frac{\Omega_m}{\Omega_{de}}\right)'=w\frac{\Omega_m}{\Omega_{de}}.
\ee This result can be viewed as a consistency proof of some of our previous results in the sense that in the course of the evolution dark energy becomes the dominant component when $w<0$, because in this case it redshifts faster than dark matter.

\begin{figure}[h!]
\begin{center}
\hspace{0.4cm}
\includegraphics[width=6.5cm,height=5cm]{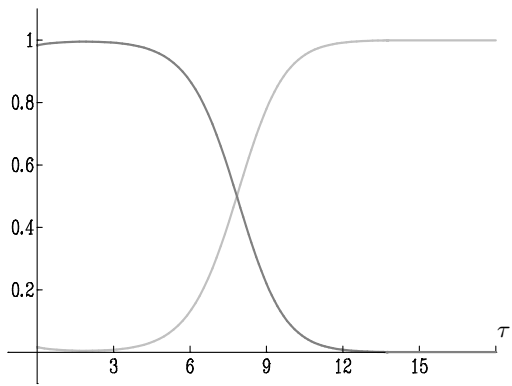}
\put(0,18){$\tau$}
\caption{Typical behavior of the fractional energy densities in the case where the phantom attractor exists. The dark and light gray lines depict $\Omega_m$ and $\Omega_{de}$ respectively.}
\label{fractions}
\end{center}
\end{figure}

Note as well that, unlike in the case of pure phantom dark energy, here the fractional energy densities of dark energy and dark matter remain both non-negligible for a long enough time span, as shown in Fig. \ref{fractions}. So at least this problem which invalidates the simplest models with a late-time phantom behavior \cite{coley} is 
absent from our model.

Needless to say that the trajectories we have presented show that for a wide range of  initial conditions for which $w\sim 1$ (close to the hyperbolas), the region $w<-1$ is reached provided $\bar m^2-\bar n^2<0$, so the crossing does indeed occur.\\

\section{Conclusions and future prospects}
The construction of possible realizations of a late-time crossing of the $w=-1$ barrier in dark energy models has become an active area. Some of the studies take advantage of the dynamical systems approach to show such crossing is possible in a family of models dubbed 
quintom cosmologies. These are configurations with two scalar fields, and in these dynamical systems investigations no other than exponential potentials have been chosen so far given some underlying physical motivation and the tractability of the choice. Here we stick to that simple election too, but consider a specification (still of exponential form) which is not a particular case of the choices made in other works. 
More specifically we have chosen a potential of the form 
$V=V_0\,e^{-\sqrt{6}(\bar m\phi+\bar n\varphi)/}$ with $V_0$, $\bar m$ and $\bar n$ positive constants, whereas (following our notation for simplicity) related papers considered either
$V=V_{01}\,e^{-2\sqrt{6}\bar m\phi}+V_{02}\,e^{-2\sqrt{6}\bar n\varphi}$ \cite{dscross1}
or 
$V=V_{01}\,e^{-2\sqrt{6}\bar m\phi}+V_{02}\,e^{-2\sqrt{6}\bar n\varphi}+
\lambda V_0\,e^{-\sqrt{6}(\bar m\phi+\bar n\varphi)}$ \cite{dscross2}, with $V_{01}$, $V_{02}$, and $\lambda$ positive constants as well.

The potential we have chosen is responsible for an interaction between the fields and leads to some surprises. We have found that, contrary to previous beliefs, quintom cosmologies may admit tracking attractors.
When the conditions for the existence of such attractors are not met, one has either phantom ($w<-1$) or de Sitter ($w=-1)$ attractors instead. This represents a departure from the situation described in \cite{dscross1} and \cite{dscross2}, because in those cases the only attractors were de Sitter-like, although the trajectories ending there had $w<-1$ transiently  before entering the de Sitter phase.

The model analyzed here can be criticized for instance on the grounds that our choice of potential is too specific, but as happens for usual quintessence models, the results obtained here might have applications for related models with more general potentials. 
Another aspect of the problem is that the coincidence problem is not solved in this model; perhaps a way to address it is to consider interactions of the scalar fields with matter as well, but this is left for future projects.

\section*{Acknowledgments}
R.L. is supported by the Spanish Ministry of Science and Education through research grants FIS2004-01626 and FIS2005-01181. G.L. has the financial support of the MES of Cuba.

\end{document}